\pdfoutput=1
\documentclass[12pt,a4]{article}
\usepackage{a4wide}
\usepackage{fullpage}
\usepackage[hidelinks]{hyperref}
\usepackage{latexsym}
\usepackage{cite}
\usepackage{mathrsfs}
\usepackage{amssymb}
\usepackage{amsmath}
\usepackage{graphicx}
\usepackage{braket}
\usepackage{enumerate}
\usepackage[usenames,dvipsnames]{xcolor}
\usepackage{todonotes}

\numberwithin{equation}{section}

\newcommand{\tr}{\mathrm{tr}}

\newcommand{\diag}{\, \mathrm{diag}}

\newcommand{\be}{\begin{equation}\label}
\newcommand{\ee}{\end{equation}}
\newcommand{\bea}{\begin{eqnarray}\label}
\newcommand{\eea}{\end{eqnarray}}

\newcommand{\beq}{\begin{eqnarray}}
\newcommand{\eeq}{\end{eqnarray}}
\newlength{\dummysp}
\newlength{\dummyssp}
\newcommand{\sspc}{\hbox{\hspace{\dummyssp}}}
\newcommand{\myref}[1]{(\ref{#1})}
\newcommand{\Ncal}{{\cal N}}
\newcommand{\nnn}{ \nonumber \\ }

\newcommand{\phib}{{\bar \phi}}
\newcommand{\cU}{{\cal U}}
\newcommand{\cUb}{{\overline{\cal U}}}

\def\({\left (}
\def\){\right )}

\date{}
\begin{document}

\begin{titlepage}

\title{3d $\mathcal{N}=4$ Super-Yang-Mills on a Lattice}
\author{Joel Giedt$^{a,}$\footnote{E-mail:  giedtj@rpi.edu} \, 
and Arthur E. Lipstein$^{b,}$\footnote{E-mail:  arthur.lipstein@durham.ac.uk} \vspace{7pt}\\ 
\normalsize \textit{ $^a$Department of Physics, Applied Physics and Astronomy} \\
\normalsize \textit{ Rensselaer Polytechnic Institute, Troy, NY 12180 USA} \\
\normalsize \textit{
$^b$Department of Mathematical Sciences}\\ \normalsize\textit{Durham University, Durham, DH1 3LE, United Kingdom}}
\maketitle
\begin{abstract}
In this paper we explore a new approach to studying three-dimensional $\mathcal{N}=4$ super-Yang-Mills on a lattice. Our strategy is to complexify the Donaldson-Witten twist of four-dimensional $\mathcal{N}=2$ super-Yang-Mills to make it amenable to a lattice formulation and we find that lattice gauge invariance forces the model to live in at most three dimensions. We analyze the renormalization of the lattice theory and show that uncomplexified three-dimensional $\mathcal{N}=4$ super-Yang-Mills can be reached in the continuum limit by supplementing the lattice action with appropriate mass terms.  

\end{abstract}

\end{titlepage}

\pagebreak
\tableofcontents

\section{Introduction}
It has been said that 4d $\Ncal=4$ super-Yang-Mills (SYM) is the hydrogen atom of quantum
field theories.  This may be motivated by the fact that it is a finite, integrable theory (in the planar limit)
about which so much is known thanks to the $SU(2,2|4)$ superconformal symmetry.  If there
is any truth in this appelation, then it is of primary interest to develop the theories
with less than the maximal amount of supersymmetry (SUSY), as these would correspond to,
say, helium or lithium atoms, which are in many ways more interesting.  In particular,
$\Ncal=2$ SYM is rich with nonpertubative phenomena such as confinement, spontaneous
chiral symmetry breaking and monopole condensation, as exhibited in the famous
Seiberg-Witten solution \cite{Seiberg:1994rs}. 

For the study of such non-perturbative questions, supersymmetric 
lattice gauge theory is potentially a very powerful approach.  
There are two known lattice formulations of 4d $\mathcal{N}=4$ SYM 
which preserve a subset of the supersymmetries:  orbifolding matrix models \cite{Kaplan:2005ta}, 
and applying geometric discretization to a topologically twisted theory \cite{Catterall:2005fd}. 
Note that lattice models derived from twisting approach are free from the fermion doubling problem \cite{Becher:1982ud}.
Of course the same will be true of the orbifold models due to their equivalence to twisted
models \cite{Catterall:2007kn}.  For further details on these two approaches, see the review \cite{Catterall:2009it}.

In this paper, we will consider analogous constructions for supersymmetric gauge theories in three dimensions, notably 3d $\mathcal{N}=4$ SYM. Since the gauge coupling has positive mass dimension in three dimensions, 3d gauge theories are asymptotically free and super-renormalizable. Hence, they can be thought of as toy models for 4d QCD. For example, many properties of high temperature QCD are captured by Euclidean 3d Yang-Mills theory coupled to scalar adjoint matter fields \cite{Appelquist:1981vg}. 

Gaiotto and Witten classified 3d $\mathcal{N}=4$ SYM theories as good, bad, and 
ugly \cite{Gaiotto:2008ak}.  For $U(N)$ gauge group and $N_f$ flavours, a good 
theory corresponds to $N_f \geq 2N$ and flows to an IR theory whose superconformal 
R-symmetry is manifest in the UV.  Ugly theories correspond to $N_f = 2N - 1$ and flow 
to an IR theory whose superconformal R-symmetry is manifest in the UV, plus a decoupled 
free sector.  Finally, bad theories correspond to $N_f \leq 2N - 2$ and do not flow to an 
IR theory whose superconformal R-symmetry is manifest in the UV, so the IR limit of 
bad theories is not as well understood.  For recent progress in this direction, see \cite{Assel:2017jgo}. 

Like 4d $\mathcal{N}=4$ SYM, the lattice formulation of 
3d $\mathcal{N}=4$ SYM can be obtained by orbifolding \cite{Cohen:2003qw} 
or topologically twisting followed by geometric discretization. As we will explain in the next section, there 
are two ways to topologically twist, one corresponding to the dimensional 
reduction of the Donaldson-Witten twist in 4d \cite{Witten:1988ze}, and the 
other known as the Blau-Thompson twist \cite{Blau:1996bx}. 
A lattice formulation based on geometric discretization of the Blau-Thompson twist was previously 
proposed in \cite{Joseph:2013jya}, so in this paper we develop a lattice formulation based on dimensionally reducing the Donaldson-Witten twist. Alternative approaches to formulating lattice 3d $\mathcal{N}=4$ SYM based on a lattice Leibnitz rule were considered in \cite{DAdda:2007hnx,Nagata:2007mz}, although these approaches have some unresolved aspects \cite{Bruckmann:2006ub,Bruckmann:2006kb}.  

Note that there are several advantages to the latter approach. 
First of all, whereas the Blau-Thompson twist utilizes an internal $SU(2)$ 
symmetry that generically becomes spontaneously broken in the IR, 
the twist we consider involves $SU(2)$ R-symmetry which is 
preserved.\footnote{We thank Stefano Cremonesi for pointing this out.}  Secondly, since our lattice model will arise from 
dimensional reduction, it will have a larger point-symmetry group 
than the lattice model arising from the Blau-Thompson twist, which is intrinsically 
three dimensional.  Indeed, we proceed by applying geometric discretization of the 
Donaldson-Witten twist of 4d $\mathcal {N}=2$ SYM, and then show that lattice 
gauge invariance only holds if the basis vectors of the lattice are linearly dependent, 
forcing the theory to live in lower dimensions.\footnote{Note that this is very similar 
to the lattice formulation of 4d $\mathcal{N}=4$ SYM, which is initially 
formulated in 5d but is then forced to live in 4d by lattice gauge invariance.}
 One disadvantage of the twisting approach compared to the orbifolding approach 
in 3d however is that in the twisted approach we must complexify the fields 
in order to implement the geometric discretization, and subsequently must 
introduce mass terms in order to decouple unwanted fields in the continuum limit.

The structure of this paper is as follows. In section \ref{continuum}, we review the 
two approaches to twisting 3d $\mathcal{N}=4$ SYM, focusing on the one we make use of this in this paperwhich is equivalent 
to dimensional reduction of the Donaldson-Witten twist of 4d $\mathcal{N}=2$ SYM), and we describe a complexification of 3d $\mathcal{N}=4$ SYM that will allow 
us to formulate the theory on a lattice. In section \ref{lattice}, we apply 
geometric descretization to the complexified theory and show that lattice 
gauge invariance forces the resulting lattice gauge theory to live in at most 
three dimensions. In section \ref{renormalization}, we discuss the renormalization 
of the lattice theory and propose mass terms to reach the desired continuum limit, 
and in section \ref{conclusion} we present our conclusions and future directions. 
We also have several appendices. In Appendix \ref{Rsymm}, we derive the 
discrete R-symmetries of the Donaldson-Witten twist, and in 
Appendix \ref{comparisonblau}, we compare the 
two twists of 3d $\mathcal{N}=4$ SYM in greater detail.


\section{Continuum Theory} \label{continuum}
Consider Euclidean 3d $\mathcal{N}=4$ SYM. The global symmetries of this theory are 
$SU(2)_E \times SU(2)_N \times SU(2)_I$, where the first 
$SU(2)$ corresponds to the rotation group in three spatial dimensions, 
the second one corresponds to an internal rotation group which arises from the 
dimensional reduction from 6d, i.e.~$ SO(6) \rightarrow SU(2)_E \times SU(2)_N $, 
and the third one is the R-symmetry group of the 6d theory \cite{Seiberg:1996nz}. 
The two known twists of this theory correspond to identifying 
$SU(2)_E$ with $SU(2)_N$ and $SU(2)_R$, respectively.  Whereas the first 
twist (constructed by Blau and Thompson \cite{Blau:1996bx}) is intrinsically 
three-dimensional, the second one can be obtained by dimensionally reducing 
the Donaldson-Witten twist of 4d $\mathcal{N}=2$ SYM \cite{Witten:1988ze}. 
A lattice theory based on the Blau-Thompson twist was previously constructed 
in \cite{Joseph:2013jya}. In this paper, we will consider the approach based 
on dimensional reduction. We relate the two approaches in Appendix \ref{comparisonblau}. 

Since we will use dimensional reduction in this paper, let us describe the 
Donaldson-Witten twist in more detail. The global symmetries of 
4d $\mathcal{N}=2$ SYM are $SU(2)_l \times SU(2)_r \times SU(2)_R \times U(1)$ 
where $SU(2)_l \times SU(2)_r$ is locally the 4d rotation group 
and $SU(2)_R \times U(1)$ is the R-symmetry group (the $U(1)$ factor arises 
from dimensional reduction from 6d and is the analogue of $SU(2)_N$ in 3d). 
The twist by $SU(2)_R$ is accomplished by identifying the twisted rotation
group, $SU(2)' \equiv \diag[ SU(2)_r \times SU(2)_R]$. 
Then the fields transform in the following 
representations of $SU(2)_l \times SU(2)' \times U(1)$:
\beq
&& \rm{bosons:}\,\,\,(1/2,1/2)^{0}+(0,0)^{2}+(0,0)^{-2}
\nnn
&& \rm{fermions:}\,\,\,(1/2,1/2)^{1}+(0,1)^{-1}+(0,0)^{-1}
\label{irrenum}
\eeq
Since $SU(2)_l \times SU(2)' \simeq SO(4)'$, the $(1/2,1/2)$
correspond to four-dimensional vector representations
of $SO(4)'$.  It will turn out that the $(0,1)$
corresponds to an antisymmetric self-dual tensor.
Thus as usual in the twisted formulations, fermions
no longer carry spinor indices, but appear as
scalars, vectors and antisymmetric tensors.
The twisted bosonic and fermionic fields enumerated in \myref{irrenum} shall subsequently 
be denoted as $\left\{ A_{\mu},\phi,\bar{\phi}\right\} $, 
and $\left\{ \psi_{\mu},\chi_{\mu\nu},\eta\right\} $, 
respectively, where $\chi_{\mu\nu}=\star\chi_{\mu\nu}\equiv\frac{1}{2}\epsilon_{\mu\nu\rho\lambda}\chi^{\rho\lambda}$, 
and $\bar{\phi}=\phi^\dagger$. Although twisted theories are usually considered in a curved background, we will be working in Euclidean flat space, so there will be no distinction between upper and lower Lorentz indices.

The Lagrangian for twisted 4d $\mathcal{N}=2$ SYM  can be written as follows:
\beq
g^{2}\mathcal{L}_{4d}^{\mathcal{N}=2} &=& \tr \bigg(\frac{1}{4}\mathcal{F}_{\mu\nu}\mathcal{F}^{\mu\nu}+\frac{1}{2}\mathcal{D}_{\mu}\bar{\phi}\mathcal{D}^{\mu}\phi-\alpha\left[\phi,\bar{\phi}\right]^{2}
\nnn && \qquad-\frac{i}{2}\eta\mathcal{D}_{\mu}\psi^{\mu}+i\alpha\phi\left\{ \eta,\eta\right\} -\frac{i}{2}\bar{\phi}\left\{ \psi_{\mu},\psi^{\mu}\right\} +\mathcal{L}_{\chi} \bigg),
\label{4dfull}
\eeq
where $\mathcal{D}_{\mu}X=\partial_{\mu}X+\mathcal{A}_{\mu}X$, $\mathcal{F}_{\mu\nu}=\left[\mathcal{D}_{\mu},\mathcal{D}_{\nu}\right]$, and
\begin{equation}
\mathcal{L}_{\chi}=tr\left(\frac{i}{8}\phi\left\{ \chi_{\mu\nu},\chi^{\mu\nu}\right\} -i\chi^{\mu\nu}\mathcal{D}_{\mu}\psi_{\nu}\right).
\label{lchi2}
\end{equation} 
Note that discrete R-symmetries fix $\alpha=\frac{1}{8}$, as we show in Appendix \ref{Rsymm}. (It is amusing that \cite{Weinberg:2000cr} also uses discrete R-symmetries to fix the form of the $\Ncal=2$ SYM
Lagrangian, albeit in the usual untwisted form; see p.~161.)
The equations of motion for $\chi$ are given by: 
\begin{equation}
\left[\chi_{\mu\nu},\phi \right]=2\left(\mathcal{D}_{[\mu}\psi_{\nu]}+*\mathcal{D}_{[\mu}\psi_{\nu]}\right).
\label{chi1eom}
\end{equation}
Plugging these equations of motion into \eqref{lchi2} then gives 
\beq
g^{2}\mathcal{L}_{4d}^{\mathcal{N}=2} &=& \tr \bigg( \frac{1}{4}\mathcal{F}_{\mu\nu}\mathcal{F}^{\mu\nu}+\frac{1}{2}\mathcal{D}_{\mu}\bar{\phi}\mathcal{D}^{\mu}\phi-\alpha\left[\phi,\bar{\phi}\right]^{2}-\frac{i}{2}\eta\mathcal{D}_{\mu}\psi^{\mu}
\nnn && \qquad -\frac{i}{2}\chi^{\mu\nu}\mathcal{D}_{\mu}\psi_{\nu}+i\alpha\phi\left\{ \eta,\eta\right\} -\frac{i}{2}\bar{\phi}\left\{ \psi_{\mu},\psi^{\mu}\right\} \bigg).
\label{4dexpand}
\eeq
This form of the Lagrangian is useful because it can be written in a way that makes a BRST symmetry manifest:
\begin{equation}
g^{2}\mathcal{L}_{4d}^{\mathcal{N}=2}=Q_{\,}tr\left[\frac{1}{4}\chi_{\mu\nu}\mathcal{F}^{\mu\nu}+\frac{1}{2}\mathcal{D}_{\mu}\bar{\phi}\psi^{\mu}+\alpha\eta\left[\phi,\bar{\phi}\right]\right]-\frac{1}{4}tr\left[\star\mathcal{F}_{\mu\nu}\mathcal{F}^{\mu\nu}\right],
\label{4dr}
\end{equation}
where $Q$ generates the following transformations \footnote{At first sight the BRST transformations may not appear to be compatible with the constraint that $\bar{\phi}$ is the complex conjugate of $\phi$, but as explained in \cite{Witten:1988ze} this is not a problem because conservation of the corresponding supercurrent is compatible with this constraint. In the next section, we will consider a complexification where $\phi$ and $\bar{\phi}$ are independent, so the BRST transformations will appear more natural.}:
\begin{equation}
Q\,\phi=0,\,\,\, Q\,\bar{\phi}=i\eta,\,\,\, Q\,\mathcal{A}_{\mu}=i\psi_{\mu},\,\,\, Q\,\eta=\left[\bar{\phi},\phi\right],\,\,\, Q\,\psi_{\mu}=\mathcal{D}_{\mu}\phi,\,\,\, Q\,\chi_{\mu\nu}=\mathcal{F}_{\mu\nu}+*\mathcal{F}_{\mu\nu}.
\label{q}
\end{equation}

Let us verify the BRST symmetry. First note that the equations of motion in \eqref{chi1eom} are invariant under the transformations in \eqref{q}. Furthermore, using equations \eqref{q} and \eqref{chi1eom} one finds that $Q^2$ generates a gauge transformation:
\begin{eqnarray*}
&& Q^{2}\phi=0,\,\,\, Q^{2}\bar{\phi}=i\left[\bar{\phi},\phi\right],\,\,\, Q^{2}\mathcal{A}_{\mu}=i\mathcal{D}_{\mu}\phi,\,\,\, Q^{2}\eta=i\left[\eta,\phi\right], 
\nnn && Q^{2}\psi_{\mu}=i\left[\psi_{\mu},\phi\right],\,\,\, Q^{2}\chi_{\mu\nu}=i\left[\chi_{\mu\nu},\phi\right].
\end{eqnarray*}   
Since $Q^2$ annihilates gauge-invariant operators, it follows that the first term in \eqref{4dr} is $Q$-exact. Moreover the second term is $Q$-closed since
\[
Q\, tr\left[\star\mathcal{F}_{\mu\nu}\mathcal{F}^{\mu\nu}\right]=4\epsilon^{\mu\nu\rho\lambda}\mathcal{D}_{\mu}\psi_{\nu}\mathcal{F}_{\rho\lambda},
\]
which vanishes after applying integration by parts and the Bianchi identity, so the twisted theory in 
\eqref{4dexpand} is indeed BRST invariant. A lattice formulation of this model was proposed in \cite{Sugino:2003yb}, although it was not based on geometric discretization. In this paper, we will employ geoemtric discretization and subsequently find that the lattice theory can live in at most three dimensions.

\subsection{Complexification} \label{complexification}
In order to make the theory amenable to geometric discretization, 
we must complexify the fields. Furthermore, when we apply geometric discretization to formulate the model on a lattice, we will find that consistency of the self-duality 
constraint of $\chi$ with lattice gauge invariance implies that the model 
can live in at most three dimensions. Taking the lattice to be three-dimensional, 
the continuum limit will then correspond to complexified 3d $\mathcal{N}=4$ SYM. 
As we will show in section \ref{mass}, the uncomplexified theory can be reached in 
the continuum limit by adding appropriate mass terms to the lattice action.

Consider the following complexification of twisted 4d $\mathcal{N}=2$ SYM: 
\beq
g^{2}\mathcal{L}^{*} &=&  \tr \bigg( \frac{1}{4}\mathcal{\mathcal{F}}_{\mu\nu}\bar{\mathcal{F}}^{\mu\nu}+\frac{1}{2}\bar{\mathcal{D}}_{\mu}\bar{\phi}\mathcal{D}_{\mu}\phi-\alpha\left[\phi,\bar{\phi}\right]^{2}
\nnn && \qquad -\frac{i}{2}\eta\bar{\mathcal{D}}_{\mu}\psi^{\mu}+i\alpha\phi\left\{ \eta,\eta\right\} -\frac{i}{2}\bar{\phi}\left\{ \psi_{\mu},\bar{\psi}^{\mu}\right\} +\mathcal{L}_{\chi} \bigg),
\label{Lcfull}
\eeq
where
\begin{equation}
\mathcal{L}_{\chi}=tr\left[\frac{i}{8}\phi\left\{ \chi_{\mu\nu},\bar{\chi}_{\mu\nu}\right\} -\frac{i}{2}\left(\chi^{\mu\nu}\mathcal{D}_{\mu}\psi_{\nu}+\bar{\chi}^{\mu\nu}\bar{\mathcal{D}}_{\mu}\bar{\psi}_{\nu}\right)\right].
\label{lchi}
\end{equation}
Note that all the fields are complex. In particular, $\bar{\mathcal{A}}=\mathcal{A}^{\dagger}$, $\bar{\eta}=\eta^{\dagger}$, $\bar{\psi}=\psi^{\dagger}$, and $\bar{\chi}=\chi^{\dagger}$, but $\bar{\phi} \neq \phi^\dagger$ so $\bar{\phi}$ and $\phi$ are independent. We define $\mathcal{D}_{\mu}X=\partial_{\mu}+\left[\mathcal{A}_{\mu},X\right]$, $\bar{\mathcal{D}}_{\mu}X=\partial_{\mu}X+\left[\bar{\mathcal{A}}_{\mu},X\right]$, $\mathcal{F}_{\mu\nu}=\left[\mathcal{D}_{\mu},\mathcal{D}_{\nu}\right]$, and $\bar{\mathcal{F}}_{\mu\nu}=\left[\bar{\mathcal{D}}_{\mu},\bar{\mathcal{D}}_{\nu}\right]$. We also impose the Hodge-duality constraint $\chi_{\mu\nu}=*\bar{\chi}_{\mu\nu}$. A similar complexification of twisted 4d $\mathcal{N}=2$ SYM was considered in \cite{Baulieu:2009aa}, where it was argued to be equivalent to twisted 4d $\mathcal{N}=4$ SYM. As mentioned above, consistency of this constraint 
with lattice gauge invariance implies that the lattice theory must be at most three-dimensional; 
details will be given the following section.

As before, we can integrate out $\chi$ to obtain the following equations of motion:
\begin{equation}
\left[\chi_{\mu\nu},\phi\right]=2\left(\bar{\mathcal{D}}_{[\mu}\bar{\psi}_{\nu]}+\star\mathcal{D}_{[\mu}\psi_{\nu]}\right).
\label{cheom}
\end{equation}
Plugging the equations of motion back into \eqref{lchi} then gives
\beq
g^{2}\mathcal{L}^{*} &=& \tr \bigg( \frac{1}{4}\mathcal{\mathcal{F}}_{\mu\nu}\bar{\mathcal{F}}^{\mu\nu}+\frac{1}{2}\bar{\mathcal{D}}_{\mu}\bar{\phi}\mathcal{D}_{\mu}\phi-\alpha\left[\phi,\bar{\phi}\right]^{2}-\frac{i}{2}\eta\bar{\mathcal{D}}_{\mu}\psi^{\mu}
\nnn && \qquad + i\alpha\phi\left\{ \eta,\eta\right\} -\frac{i}{2}\bar{\phi}\left\{ \psi_{\mu},\bar{\psi}^{\mu}\right\} -\frac{i}{2}\chi^{\mu\nu}\mathcal{D}_{\mu}\psi_{\nu} \bigg) ,
\label{4dcomplex}
\eeq
which can be written in a manifestly BRST-invariant form as follows: 
\begin{equation}
g^{2}\mathcal{L}^{*}=Q_{\,}tr\left[\frac{1}{4}\chi_{\mu\nu}\mathcal{F}^{\mu\nu}+\frac{1}{2}\bar{\mathcal{D}}_{\mu}\bar{\phi}\psi^{\mu}+\alpha\eta\left[\phi,\bar{\phi}\right]\right]-\frac{1}{4}tr\left(\star\mathcal{F}_{\mu\nu}\mathcal{F}^{\mu\nu}\right),
\label{4dc}
\end{equation}
where $Q$ generates the following transformations:
\[
Q\,\phi=0,\,\,\, Q\,\bar{\phi}=i\eta,\,\,\, Q\,\mathcal{A}_{\mu}=i\psi_{\mu},\,\,\, Q\,\bar{\mathcal{A}}_{\mu}=i\bar{\psi}_{\mu},
\]
\begin{equation}
Q\,\eta=\left[\bar{\phi},\phi\right],\,\,\, Q\,\psi_{\mu}=\mathcal{D}_{\mu}\phi,\,\,\, Q\,\bar{\psi}_{\mu}=\bar{\mathcal{D}}_{\mu}\phi,\,\,\, Q\,\chi_{\mu\nu}=\bar{\mathcal{F}}_{\mu\nu}+*\mathcal{F}_{\mu\nu}.
\label{Qco}
\end{equation}
Note that that the equations of motion in \eqref{cheom} are invariant under the transformations in \eqref{Qco}. Using \eqref{Qco} and \eqref{cheom} it is not difficult to show that $Q$ squares into a gauge transformation implying that the first term in \eqref{4dc} is BRST exact. Moreover the second term is BRST closed since acting with $Q$ gives a term that vanishes by the Bianchi identity, as shown in the previous subsection. Hence the complexified theory in \eqref{Lcfull} is BRST invariant. Our next task will be to formulate this model on a lattice in such a way that the BRST symmetry is preserved and becomes enhanced to $\mathcal{N}=2$ supersymmetry in the continuum limit.    

\section{Lattice Theory} \label{lattice}
In this section, we will generalize the complexified Lagrangian in \eqref{Lcfull} to a lattice. 
The fields $\phi(n),\bar{\phi}(n),\eta(n)$ will reside on the lattice site 
$n = \sum_{\mu=1}^d n_\mu e_\mu$, $n_\mu \in \mathbb{Z}$, 
while the gauge field $\mathcal{A}_{\mu}(n)$ will be generalized
to a Wilson line $\mathcal{U}_{\mu}(n)$ residing on the link $\left(n,n+e_{\mu}\right)$
as will $\psi_{\mu}(n)$, and the field $\chi_{\mu\nu}(n)$ will reside
on the link $\left(n+e_{\mu}+e_{\nu},n\right)$. 
In these expressions, $e_\mu$ are principle vectors of the
lattice, which will turn out to be linearly dependent,
and are proportional to the lattice spacing.
Under a lattice gauge transformation, the fields transform as follows: 
\beq
\left\{ \phi(n),\bar{\phi}(n),\eta(n)\right\} &\rightarrow& G(n)\left\{ \phi(n),\bar{\phi}(n),\eta(n)\right\} G^{\dagger}(n)
\nnn
\left\{ \mathcal{U}_{\mu}(n),\psi_{\mu}(n)\right\} &\rightarrow& G(n)\left\{ \mathcal{U}_{\mu}(n),\psi_{\mu}(n)\right\} G^{\dagger}\left(n+e_{\mu}\right)
\nnn
\chi_{\mu\nu}(n) &\rightarrow& G\left(n+e_{\mu}+e_{\nu}\right)\chi_{\mu\nu}(n)G^{\dagger}(n).
\label{gauge}
\eeq
Here, $G(n)$ are elements of the gauge group of the target (continuum) theory.
As is usual with the twisted (or orbifold) formulations, the gauge group
must be $U(N)$, in order to be consistent with the scalar supersymmetry
algebra, \eqref{scalarsusy} below.  In particular, the relation
$Q\,\mathcal{U}_{\mu}(n)=i\psi_{\mu}(n)$ must contain $GL(N,\mathbb{C})$
valued fields on both sides, because $\mathcal{U}_{\mu}(n) = 1 + a \mathcal{A}_{\mu}(n) + \cdots$
in the continuum limit, so $\psi_{\mu}(n)$ must also have an expansion
that includes the unit matrix.
The lattice covariant derivatives are defined to be
\[
\mathcal{D}_{\mu}^{+}f(n)=\mathcal{U}_{\mu}(n)f\left(n+e_{\mu}\right)-f(n)\mathcal{U}_{\mu}(n)
\]
\[
\mathcal{D}_{\mu}^{+}f_{\nu}(n)=\mathcal{U}_{\mu}(n)f_{\nu}\left(n+e_{\mu}\right)-f_{\nu}(n)\mathcal{U}_{\mu}\left(n+e_{\nu}\right)
\]
\[
\bar{\mathcal{D}}_{\mu}^{-}f_{\mu}(n)=f_{\mu}(n)\bar{\mathcal{U}}_{\mu}(n)-\bar{\mathcal{U}}_{\mu}\left(n-e_{\mu}\right)f_{\mu}\left(n-e_{\mu}\right)
\]
\[
\bar{\mathcal{D}}_{\mu}^{-}f_{\nu\lambda}(n)=f_{\nu\lambda}(n)\bar{\mathcal{U}}_{\mu}\left(n-e_{\mu}\right)-\bar{\mathcal{U}}_{\mu}\left(n+e_{\nu}+e_{\lambda}-e_{\mu}\right)f_{\nu\lambda}\left(n-e_{\mu}\right).
\]
In terms of these lattice derivatives, the gauge field strength is then given by
\[
\mathcal{F}_{\mu\nu}(n)=\mathcal{D}_{\mu}^{+}\mathcal{U}_{\nu}(n).
\]
Similarly, we define the complex conjugate lattice derivatives as follows:
\[
\mathcal{\bar{D}}_{\mu}^{+}f(n)=f\left(n+e_{\mu}\right)\bar{\mathcal{U}}_{\mu}(n)-\mathcal{\bar{U}}_{\mu}(n)f(n)
\]
\[
\mathcal{\bar{D}}_{\mu}^{+}f_{\nu}(n)=f_{\nu}\left(n+e_{\mu}\right)\mathcal{\bar{U}}_{\mu}(n)-\bar{\mathcal{U}}_{\mu}\left(n+e_{\nu}\right)f_{\nu}(n)
\]
\[
\mathcal{D}_{\mu}^{-}f_{\mu}(n)=\mathcal{U}_{\mu}(n)f_{\mu}(n)-f_{\mu}\left(n-e_{\mu}\right)\mathcal{U}_{\mu}\left(n-e_{\mu}\right)
\]
\[
\mathcal{D}_{\mu}^{-}f_{\nu\lambda}(n)=\mathcal{U}_{\mu}\left(n-e_{\mu}\right)f_{\nu\lambda}(n)-f_{\nu\lambda}\left(n-e_{\mu}\right)\mathcal{U}_{\mu}\left(n+e_{\nu}+e_{\lambda}-e_{\mu}\right),
\]
in terms of which the complex conjugate field strength is given by
\[
\bar{\mathcal{F}}_{\mu\nu}(n)=-\bar{\mathcal{D}}_{\mu}^{+}\bar{\mathcal{U}}_{\nu}(n).
\] 

For the complexified theory in \eqref{Lcfull}, supersymmetry requires that $\chi_{\mu\nu}=*\bar{\chi}_{\mu\nu}$ in order to have an equal number of bosons and fermions. Let us therefore impose the following analogous constraint on the lattice fields $\chi$ and $\bar{\chi}$:
\begin{equation}
\chi_{\mu\nu}(n)=\frac{1}{2}\epsilon_{\mu\nu\rho\lambda}\bar{\chi}_{\rho\lambda}\left(n+e_{\mu}+e_{\nu}\right).
\label{eq:sdconstraint}
\end{equation}
If we apply the lattice gauge transformations in \eqref{gauge}, we find that the above equation is left invariant if and only if the basis vectors of the lattice are linearly dependent:
\begin{equation}
\sum_{\mu=1}^{4}e_{\mu}=0.
\label{eq:esum}
\end{equation}
Note that the constraint in \eqref{eq:sdconstraint} is the unique choice which reduces to $\chi_{\mu\nu}=*\bar{\chi}_{\mu\nu}$ in the continuum limit and respects lattice gauge and $Q$-invariance. To see this, consider a more general ansatz:
\[
\chi_{\mu\nu}(n)=\frac{1}{2}\epsilon_{\mu\nu\rho\lambda}\bar{\chi}_{\rho\lambda}\left(n+\Delta \right).
\]
Applying the gauge transformations in \eqref{gauge} then implies the constraints $\Delta=e_{\mu}+e_{\nu}$ and $\Delta+e_{\rho}+e_{\lambda}=0$, from which \eqref{eq:sdconstraint} and \eqref{eq:esum} follow. One could consider generalizing this ansatz by dressing it with link variables $ \left(\mathcal{U}_{\mu},\bar{\mathcal{U}}_{\mu}\right)$, but this will spoil $Q$-inavariance of the lattice theory since the link variables must transform non-trivially under BRST transformations, as we will see shortly. One could also consider replacing $\bar{\chi}$ with $\chi$ on the right-hand-side of the above ansatz since this would imply the same reduction in the number of femrions, but in that case one finds that there is no choice of $\Delta$ consistent with the lattice gauge transformations in \eqref{gauge}.   

Hence, the lattice can be at most three-dimensional. In this case, the complexifed Lagrangian in \eqref{Lcfull} has the following lattice generalization:
\[
\mathcal{L}={\rm tr}\left(\frac{1}{4}\bar{\mathcal{F}}_{\mu\nu}(n)\mathcal{F}_{\mu\nu}(n)+\frac{1}{2}\bar{\mathcal{D}}_{\mu}^{+}\bar{\phi}(n)\mathcal{D}_{\mu}^{+}\phi(n)-\alpha\left[\phi(n),\bar{\phi}(n)\right]^{2}\right.
\]
\begin{equation}
\left.+\frac{i}{2}\bar{\mathcal{D}}_{\mu}^{+}\eta(n)\psi_{\mu}(n)+i\alpha\phi(n)\left\{ \eta(n),\eta(n)\right\} -\frac{i}{2}\bar{\phi}(n)\left(\psi_{\mu}(n)\bar{\psi}_{\mu}(n)+\bar{\psi}_{\mu}\left(n-e_{\mu}\right)\psi_\mu\left(n-e_{\mu}\right)\right)\right) +\mathcal{L}_{\chi},
\label{lattfull}
\end{equation}
where
\[
\mathcal{L}_{\chi}={\rm tr}\left[\frac{i}{8}\left(\phi(n)\bar{\chi}_{\mu\nu}(n)\chi_{\mu\nu}(n)+\phi\left(n+e_{\mu}+e_{\nu}\right)\chi_{\mu\nu}(n)\bar{\chi}_{\mu\nu}(n)\right)\right.
\]
\begin{equation}
\left.-\frac{i}{2}\left(\bar{\chi}_{\mu\nu}(n)\bar{\mathcal{D}}_{\mu}^{+}\bar{\psi}_{\nu}(n)+\chi_{\mu\nu}(n)\mathcal{D}_{\mu}^{+}\psi_{\nu}(n)\right)\right].
\label{lc2}
\end{equation}
Using the constraints in \eqref{eq:sdconstraint} and \eqref{eq:esum}, one obtains the following equations of motion for $\bar{\chi}$:
\begin{equation}
2\left(\bar{\mathcal{D}}_{[\mu}^{+}\bar{\psi}_{\nu]}(n)+\frac{1}{2}\epsilon_{\mu\nu\rho\lambda}\mathcal{D}_{\rho}^{+}\psi_{\lambda}\left(n+e_{\mu}+e_{\nu}\right)\right)=\chi_{\mu\nu}(n)\phi(n)-\phi\left(n+e_{\mu}+e_{\nu}\right)\chi_{\mu\nu}(n).
\label{chieomlatt}
\end{equation}
Plugging the equations of motion back into \eqref{lc2} then gives
\begin{equation}
\mathcal{L}_{\chi}=-\frac{i}{4}{\rm tr}\left(\bar{\chi}_{\mu\nu}(n)\bar{\mathcal{D}}_{\mu}^{+}\bar{\psi}_{\nu}(n)+\chi_{\mu\nu}(n)\mathcal{D}_{\mu}^{+}\psi_{\nu}(n)\right).
\label{lchi1}
\end{equation}
The Lagrangian can subsequently be expressed in a manifestly BRST invariant form as follows:
\[
\mathcal{L}= Q\,{\rm tr}\left(\frac{1}{4}\chi_{\mu\nu}(n)\mathcal{F}_{\mu\nu}(n)+\frac{1}{2}\bar{\mathcal{D}}_{\mu}^{+}\bar{\phi}(n)\psi_{\mu}(n)+\alpha\eta(n)\left[\phi(n),\bar{\phi}(n)\right]\right)
\]
\begin{equation}
-\frac{1}{8}\epsilon_{\mu\nu\rho\lambda}{\rm tr}\left(\mathcal{F}_{\mu\nu}(n)\mathcal{F}_{\rho\lambda}\left(n+e_{\mu}+e_{\nu}\right)\right),
\label{3dlatt}
\end{equation}
where the BRST operator $Q$ acts according to
\beq
&& Q\,\phi(n)=0,\,\,\, Q\,\bar{\phi}(n)=i\eta(n),
\nnn
&&
Q\, \eta(n)=\left[\bar{\phi}(n),\phi(n)\right],\,\,\,
\nnn
&&
Q\,\mathcal{U}_{\mu}(n)=i\psi_{\mu}(n),\,\,\, Q\,\bar{\mathcal{U}}_{\mu}(n)=-i\bar{\psi}_{\mu}(n)
\nnn
&&
Q\,\psi_{\mu}(n)=\mathcal{D}_{\mu}^{+}\phi(n),\,\,\, Q\,\bar{\psi}_{\mu}(n)=\bar{\mathcal{D}}_{\mu}^{+}\phi(n)
\nnn
&&
Q\,\chi_{\mu\nu}(n)=\bar{\mathcal{F}}_{\mu\nu}(n)+\frac{1}{2}\epsilon_{\mu\nu\rho\lambda}\mathcal{F}_{\rho\lambda}\left(n+e_{\mu}+e_{\nu}\right).
\label{scalarsusy}
\eeq

To verify that the lattice theory is BRST-invariant, first note that the constraint in \eqref{eq:sdconstraint} and the equations of motion in \eqref{chieomlatt} are invariant under the tranformations in \eqref{scalarsusy}, and the second term in \eqref{3dlatt} is $Q$-closed since
\[
Q\,\sum_{n}\epsilon_{\mu\nu\rho\lambda}{\rm tr}\left[\mathcal{F}_{\mu\nu}(n)\mathcal{F}_{\rho\lambda}\left(n+e_{\mu}+e_{\nu}\right)\right]=-4i\sum_{n}\epsilon_{\mu\nu\rho\lambda}\psi_{\nu}\left(n-e_{\nu}\right)\mathcal{D}_{\mu}^{-}\mathcal{F}_{\rho\lambda}\left(n+e_{\mu}\right)=0,
\]
where we used \eqref{eq:esum} and the lattice Bianchi identity $\epsilon_{\mu\nu\rho\lambda}\mathcal{D}_{\mu}^{-}\mathcal{F}_{\rho\lambda}\left(n+e_{\mu}\right)=0$. Furthermore, using \eqref{scalarsusy} and \eqref{chieomlatt}, one finds that  $Q^2$ generates a lattice gauge transformation:
\[
Q^{2}\phi(n)=0,\,\,\, Q^{2}\bar{\phi}(n)=i\left[\bar{\phi}(n),\phi(n)\right],
\]
\[
Q^{2}\eta(n)=i\left[\eta(n),\phi(n)\right],\,\,\, 
\]
\[
Q^{2}\mathcal{U}_{\mu}(n)=i\mathcal{D}_{\mu}^{+}\phi(n),\,\,\, Q^{2}\bar{\mathcal{U}}_{\mu}(n)=-i\bar{\mathcal{D}}_{\mu}^{+}\phi(n).
\]
\[
Q^{2}\psi_{\mu}(n)=i\left(\psi_{\mu}(n)\phi\left(n+e_{\mu}\right)-\phi(n)\psi_{\mu}(n)\right)
\]
\[
Q^{2}\bar{\psi}_{\mu}(n)=i\left(\bar{\psi}_{\mu}(n)\phi\left(n\right)-\phi\left(n+e_{\mu}\right)\bar{\psi}_{\mu}(n)\right).
\]
\[
Q^{2}\chi_{\mu\nu}(n)=2i\left(\bar{\mathcal{D}}_{[\mu}^{+}\bar{\psi}_{\nu]}(n)+\frac{1}{2}\epsilon_{\mu\nu\rho\lambda}\mathcal{D}_{\rho}^{+}\psi_{\lambda}\left(n+e_{\mu}+e_{\nu}\right)\right).
\]
Hence the first term in \eqref{3dlatt} is $Q$-exact and the lattice theory in \eqref{lattfull} is indeed BRST-invariant.  

\subsection{Base Space}
To avoid unwanted renormalizations, it is advantageous to have maximal point group symmetry.
Thus we would like to have principal vectors $e_1, e_2, e_3$ and $e_4$, arranged such that
the symmetry is just $S_4$, the permutation group of four numbers, $(1234) \to (2134)$, etc. This is achieved with the $A_3^*$ lattice \cite{Kaplan:2005ta,Saidi:2014wra}, which has the hyper-triangular constraints
\beq
e_\mu \cdot e_\nu = \delta_{\mu \nu} - \frac{1}{4}, \quad
e_1+e_2+e_3+e_4=0, \quad \sum_{\mu=1}^4 e_\mu^{\sspc i} e_\nu^{\sspc j} = \delta^{ij},
\label{basisrules}
\eeq
where $i,j=1,2,3$ label the components of the four vectors. An explicit choice of the principal
vectors is
\beq
&& e_1 = \left( \sqrt{\frac{3}{4}}, 0, 0 \right), \quad
e_2= \left( - \frac{1}{\sqrt{12}}, \sqrt{\frac{2}{3}}, 0 \right) \nonumber \\
&& e_3 = \left( - \frac{1}{\sqrt{12}}, - \frac{1}{\sqrt{6}}, \frac{1}{\sqrt{2}} \right),
\quad e_4 = \left( - \frac{1}{\sqrt{12}}, - \frac{1}{\sqrt{6}}, -\frac{1}{\sqrt{2}} \right)
\eeq
It can be checked that these satisfy \myref{basisrules}.

The lattice $\Lambda$ can be specified by
\beq
\Lambda = \{ \; n_1 e_1 + n_2 e_2 + n_3 e_3 \; | \; n_i \in \mathbb{Z} \; \forall \; i=1,2,3 \; \}
\eeq
The vector ${\bf n}=(n_1,n_2,n_3)$ is then associated with a site
on an {\it abstract} cubic lattice, which is how a computer code
would ``see'' the lattice.  In particular, periodic boundary
conditions would be imposed {\it via} $n_i \simeq n_i + L_i$
with $L_i$ the size of the abstract torus in the direction $i$. Note that the integer-valued vectors labelling the sites of abstract are related to the R-charges defined in the orbifold
formulation \cite{Kaplan:2005ta,Cohen:2003qw}.

The four basic directions $e_\mu$ are to be associated with the link fields
${\cal A}_\mu$ and $\psi_\mu$.  Note that the $S_4$ point-symmetry group is a subgroup of the twisted rotation group $SU(2)'$ and the lattice fields transform in reducible representations of the point symmetry group. For example, the four-vector $A_\mu$ decomposes into a three-vector and a scalar, which corresponds to the $S_4$ symmetric linear combination $\sigma = \sum_{\mu=1}^4 A_\mu$. The parts of $A_\mu$
that are orthogonal to this symmetric combination are the {\it bona fide}
gauge fields of the 3d theory. Morover if we add mass terms for the imaginary parts of the fields (as we describe in section \ref{mass}), in continuum limit we will be left with one real scalar and one gauge field along with two real scalars coming from $(\phi, \bar{\phi})$, which is precisely the bosonic field content of 3d $\mathcal{N}=4$ SYM. For more details of the dimensional reduction, see Appendix \ref{comparisonblau}.

\section{Renormalization} \label{renormalization}
To understand the renormalization of the lattice
theory, first we have to outline the dimensions
of fields and parameters.  Note that, in terms
of mass dimensions, 
\beq
\left[ \frac{1}{g^2} \right] = -1 \quad \Rightarrow \quad [g] = 1/2
\eeq
This can be understood in terms of dimensional reduction, which
yields
\beq
\frac{1}{g^2} = \frac{L_4}{g_\text{4d}^2}
\eeq
where $L_4$ is the size of the fourth dimension that is reduced out,
and of course the 4d gauge coupling is dimensionless.
It follows that, as written, all fields have the same dimensions
as in 4d.  However, this is not convenient for an analysis of
relevant/marginal/irrelevant operator classification, so we
scale out the coupling to get canonical kinetic terms before
proceeding.  For instance,
\beq
{\cal A}_\mu \to g {\cal A}_\mu, \quad {\cal F}_{\mu\nu} \to g {\cal F}_{\mu\nu}, \quad
\phi \to g \phi, \quad {\bar \phi} \to g {\bar \phi}, \quad \psi_\mu \to g \psi_\mu, \quad \eta \to g \eta \ldots
\label{mrs}
\eeq
Hence after the rescaling,
\beq
[{\cal A}_\mu ] = 1/2, \quad [{\cal F}_{\mu\nu} ] = 3/2, \quad  [ \phi ]=[ {\bar \phi} ] =1/2 , \quad
[\psi_\mu]=[\eta]=1 ....
\eeq
Notice that the scalar potential term becomes $g^2 [ \phi , {\bar \phi} ]^2$, so that the
operator has mass dimension 2 and is {\it relevant.}  This is very important for understanding the unwanted
radiative corrections that occur when we flow to the long distance
effective theory (given that the lattice regulator breaks SUSY explicitly
and allows for non-SUSY renormalizations).

To further understand these matters with canonical normalization,
notice the SUSY relation $Q {\cal A}_\mu = i \psi_\mu$ implies
that the supercharge carries mass dimension, $[Q] = 1/2$.  This will allow
us to analyze the renormalizations from the point of view of
$Q$-exact terms.  Notice also that the SUSY variation of $\eta$ becomes
$Q \eta = g [{\bar \phi}, \phi]$.  Thus we see also from this perspective that
$Q ( \eta [\phi, {\bar \phi}] )$ is a relevant operator, with mass dimension
5/2, though the $[ {\bar \phi}, \phi ]^2$ part of it should really be counted as mass
dimension 2, because of the appearance of $g$, for the purposes
of RG analysis.  In detail, taking into account the rescalings \myref{mrs},
\beq
\frac{1}{g^2} Q \tr \eta [\phi, {\bar \phi}] \to g Q \tr \eta [\phi, {\bar \phi}]
= g^2 \tr [\phi, {\bar \phi}]^2 + i g \tr \eta [ \phi, \eta ]
\eeq
so we see that we have both dimension $2$ and $5/2$ operators, with
the additional dimensions (to get to $3$) soaked up by powers of $g$,
which has dimension $1/2$.  

Because in 3d the coupling carries mass dimension, renormalizations are highly constrained.
If in the long distance effective theory a marginal operator is generated, it must
enter at one loop as ${\cal O}(g^2 a) $, and with higher powers of $g^2 a$ at higher
orders.  Here, it is important that we keep track of the
dimensionless quantity $g^2 a$, because the coefficient
of a marginal operator must be dimensionless, yet loop
corrections (with canonical normalization of kinetic terms)
will be powers of $g^2$.
 Thus in the continuum limit $a \to 0$, all of these radiative corrections
vanish.  For this reason we do not have to worry about marginal operators in terms
of recovering SUSY.  Irrelevant operators simply come with additional powers
of the lattice spacing $a$, so these are also not troublesome.  Thus the
operators that we must focus our attention on are the relevant operators,
which in 3d with canonical kinetic terms are operators with $d_{\cal O} < 3$.
In the remainder of this subsection we enumerate such operators allowed by the lattice
symmetries, which will give us a count on the number of fine-tunings
required to achieve the SUSY continuum limit.
Thus we set aside a
rather long list of marginal operators that are consistent with $Q$ and $S_4$
invariance, such as:
\beq
Q \tr {\cal D}_\mu {\bar \phi} \psi_\mu, \quad Q \tr ( \sum_\mu {\cal D}_\mu \phi ) \eta,
\quad Q \tr ( \eta \sum_{\mu < \nu} {\cal F}_{\mu \nu} ), \quad \ldots
\eeq

Firstly, as noted above, $Q \tr ( \eta [ \phi, {\bar \phi}] )$ is a relevant operator.  In particular, the
part of it that gives $[ \phi, {\bar \phi} ]^2$ could appear at one loop with $g^2$ and
radiative effects will allow this to have a dimensionless coefficient
that is different from the SUSY theory.  So from this we get one
fine-tuning, but it is not a new operator that we have to add to
the lattice theory to achieve the SUSY long distance limit.  In particular,
for relatively coarse lattice spacings and weak $g\sqrt{a}$, we do not
anticipate large deviations from SUSY occuring from this operator, based
on experience in the 4d ${\cal N}=4$ lattice theory. 
 With these sorts of considerations we finally find
the following list of relevant operators that are allowed by lattice gauge invariance,
Q symmetry, and the $S_4$ point group symmetry:
\beq
&& Q \tr ( \eta [ \phi, {\bar \phi}] ), \quad Q \tr ( \eta \{ \phi, {\bar \phi} \} ), \quad
Q \tr ( \eta \phi ), \quad Q \tr ( \eta {\bar \phi} ), 
\nnn && Q ( \tr \eta  \tr \{ \phi, {\bar \phi} \} ), \quad
Q ( \tr \eta \tr \phi ), \quad Q ( \tr \eta \tr {\bar \phi} )
\label{rts}
\eeq
The double trace operators, that involve the trace of a single field, are possible
because the group is $U(N)$.  Note however that, in analogy to the 4d $\Ncal=4$ theory,
there is a fermionic shift symmetry in the bare theory,
\beq
\eta \to \eta + b \mathbb{I}
\eeq
where $b$ is a constant Grassmann number and $\mathbb{I}$ is an $N \times N$ unit matrix.
This forbids all of the terms in \myref{rts} individually, except for $Q \tr ( \eta [ \phi, {\bar \phi}] )$,
which already appears in the tree action.  On the other hand, again in analogy to the
4d $\Ncal=4$ theory, there are three linear combinations of the other terms that are allowed by
the shift symmetry:
\beq
&& Q \tr ( \eta \{ \phi, {\bar \phi} \} ) - \frac{1}{N} Q ( \tr \eta  \tr \{ \phi, {\bar \phi} \} ) \nnn
&& Q \tr ( \eta \phi ) - \frac{1}{N} Q ( \tr \eta \tr \phi ) \nnn
&& Q \tr ( \eta {\bar \phi} ) - \frac{1}{N} Q ( \tr \eta \tr {\bar \phi} )
\eeq
These three {\it new} operators are allowed by all the lattice symmetries.  They generate terms that
are cubic and quartic in scalar fields, as well as a mass term for the $SU(N)$ part of $\eta$.
E.g.,
\beq
Q \tr ( \eta {\bar \phi} ) - \frac{1}{N} Q ( \tr \eta \tr {\bar \phi} ) = \frac{g}{N}{\rm tr}\left[\phi,\bar{\phi}\right]{\rm tr}\bar{\phi}
\eeq 

In each of the cases above, we end up with a contribution from dimension $2$ operators after
applying $Q$.  Loop corrections beyond $g^2$ would necessarily come with positive powers
of the lattice spacing $a$, and therefore do not appear in the continuum limit.
The result of this is that all of the quantum corrections that have to be cancelled
by counterterms appear at $1$-loop.  This renders the fine-tuning of the lattice
theory quite manageable.

\subsection{Mass Terms} 
\label{mass}
A first, basic requirement of the lattice formulation is that it reproduce the desired continuum
limit, notably ${\cal N}=4$ SYM in 3d. As for lattice 4d $\mathcal{N}=4$ SYM, 
we must add the following mass terms to ensure 
that the gauge link fields have the form 
$\mathcal{U}_{\mu}(n) = 1 + a \mathcal{A}_{\mu}(n) + \cdots$ in the continuum limit:
\beq
\mathcal{L}_U = m_U^2 \left( \frac{1}{N} \sum_\mu \tr \left( {\cal U}_\mu(n) \overline{\cal U}_\mu (n) \right)- 1 \right)^{2}.
\label{lfpeq}
\eeq 
Let us denote the Hermitian part of $\mathcal{A}_{\mu}$ as $B_{\mu}$. In Appendix \ref{lfp}, we show that \eqref{lfpeq}   provides a mass term for the abelian part of $B$, which protects the dynamical lattice spacing $a$ from uncontrolled fluctuations. 

Because of the complexification that was introduced above, the continuum theory has double the desired spectrum, so we must add additional mass terms to decouple the unwanted fields. Note that abelian part of the $B$ field is already decoupled by the mass terms in \eqref{lfpeq}. To decouple the non-abelian part of $B$, we add the following additional mass terms:
\begin{equation}
\mathcal{L}_{B}=m_{B}^{2}\sum_{\mu}\tr\left[\left({\cal U}_{\mu}(n)\overline{{\cal U}}_{\mu}(n)-1/N \tr\left({\cal U}_{\mu}(n)\overline{{\cal U}}_{\mu}(n)\right)\right)^{2}\right].
\label{Bmass}
\end{equation}
For the scalar fields $\phi$ and $\bar{\phi}$, we take the mass terms to be
\begin{equation}
\mathcal{L}_{\phi} = m_{\phi}^2 \tr \left| \phi(n)^{\dagger}-\bar{\phi}(n)\right |^{2}. 
\label{phimass}
\end{equation}
If we break up the scalar fields into real components
\beq
\phi = \phi_R + i \phi_I, \quad
\phib = \phib_R + i \phib_I,
\eeq
the mass term then takes the form
\beq
(\phi_R, \phi_I, \phib_R, \phib_I)
\begin{pmatrix}
1 & 0 & -1 & 0 \cr
0 & 1 & 0 & 1 \cr
-1 & 0 & 1 & 0 \cr
0 & 1 & 0 & 1 
\end{pmatrix}
\begin{pmatrix}
\phi_R \cr \phi_I \cr \phib_R \cr \phib_I
\end{pmatrix}.
\eeq
This matrix has two eigenvectors with eigenvalue $2$:
\beq
\phi_I^+ = \phi_I + \phib_I, \quad \phi_R^- = \phi_R - \phib_R
\eeq
and two eigenvectors with eigenvalue $0$ (i.e., states that will survive in the
low energy spectrum):
\beq
\phi_I^- =\phi_I - \phib_I, \quad  \phi_R^+ = \phi_R + \phib_R.
\eeq

Unlike the scalars, the real and imaginary parts of the fermionic fields $\psi$ and $\chi$ do not transform covariantly under lattice gauge transformations.
On the other hand, the following fermionic fields do transform covariantly
and reduce to their real and imaginary parts in the continuum limit:
\beq
\psi_{\mu}^{\pm}(n) &=& \psi_{\mu}(n)\pm U_{\mu}(n)\bar{\psi}_{\mu}(n)U_{\mu}(n)
\nnn
\chi_{\mu\nu}^{\pm}(n)&=&\chi_{\mu\nu}(n)\pm\bar{U}_{\mu}(n+e_{\nu})\bar{U}_{\nu}(n)\bar{\chi}_{\mu\nu}(n)\bar{U}_{\mu}(n+e_{\nu})\bar{U}_{\nu}(n),
\label{aproj}
\eeq
where repeated indices aren't summed. We also define
\[
\eta^{\pm}(n)=\eta(n)\pm\bar{\eta}(n).
\]
Note that $\bar{\eta}^{\pm}(n)=\pm\eta(n)$. Also note that the Hodge-duality constraint on $\chi$ translates into
the following constraint relating light fields to heavy fields:
\[
\chi_{\mu\nu}^{+}(n)-\frac{1}{2}\epsilon_{\mu\nu\rho\lambda}\bar{\chi}_{\rho\lambda}^{+}(n+e_{\mu}+e_{_{\nu}})=\frac{1}{2}\epsilon_{\mu\nu\rho\lambda}\bar{\chi}_{\rho\lambda}^{-}(n+e_{\mu}+e_{_{\nu}})-\chi_{\mu\nu}^{-}(n).
\]
We then define the following fermionic mass terms:
\begin{equation}
\mathcal{L}_{\Psi}=m_{\Psi}\left[\bar{\psi}_{\mu}^{-}(n)\psi_{\mu}^{-}(n)+\bar{\chi}_{\mu\nu}^{-}(n)\chi_{\mu\nu}^{-}(n)+\eta^{-}(n)\left(\psi_{\mu}^{-}(n)\bar{U}_{\mu}(n)+U_{\mu}(n)U_{\nu}(n+e_{\nu})\chi_{\mu\nu}^{-}(n)+c.c.\right)\right],
\label{fermimass}
\end{equation}
where repeated indices are now summed over. 

After writing the Lagrnagian in terms of the light and heavy matter fields defined above, one can integrate out the heavy fields 
at nonzero lattice spacing to obtain a gauge-invariant Lagrangian in terms of light matter fields. Although it is not necessary to obtain the correct continuum limit, we may also take the real part of the Lagrangian in \eqref{lattfull} as this will ensure that the action is real at non-zero lattice spacing, which is more convenient for numerical calculations. It may also be interesting to simulate the complex action using recently developed Lefschetz thimble and complex Langevin techniques (see for example \cite{Cristoforetti:2012su}).  

All of the mass parameters in \eqref{Bmass}, \eqref{phimass}, and \eqref{fermimass} are of the same order, large
compared to the dynamical scale $\Lambda_{dyn}$ of 3d $\Ncal=4$ SYM. Since the mass terms cannot be obtained from  $Q$-exact terms, they will violate this symmetry. Note that taking the real part of the Lagrangian in \eqref{lattfull} also breaks lattice susy \footnote{We thank Loganayagam R and Masanori Hanada for discussions on this point.}. On the other hand, since the model is super-renormalizable this will just introduce a finite number of additional counter-terms up to two loops so renormalizing the theory will still be a managable task. Since the unwanted fields are only coupled to the $\Ncal=4$ SYM sector through gauge interactions and scalar interactions, they decouple according to the Appelquist and Carazzone theorem \cite{Appelquist:1974tg} and we are therefore left with the 3d $\Ncal=4$ field content at low energies. Because we take $m_i \gg \Lambda_{dyn}$, where $\Lambda_{dyn} \sim g^2$ is the scale of the target 3d gauge theory, the effective gauge coupling $g^2/m_i$ is weak at the
scale where they decouple, and this perturbative analysis of decoupling is reliable. We leave a detailed analysis of perturbative renormalization for future work.

\section{Conclusion} \label{conclusion}

In this paper, we explore a new approach to formulating 3d $\mathcal{N}=4$ SYM on a lattice. Starting with a complexifcation of the Donaldson-Witten twist of 4d $\mathcal{N}=2$ SYM, we apply geometric discretization and find that lattice gauge invariance is only consistent with a certain Hodge-duality constraint on the lattice fermions if the basis vectors of the lattice are linearly dependent, implying that the model can live in at most three dimensions. Choosing the basis vectors to form a tetrahedron (or equivalently to span an $A_3^*$ lattice), the resulting lattice gauge theory has an $S_4$ point symmetry, in contrast to the lattice formulation based on the Blau-Thompson twist which has an $S_3$ point symmetry group. We then analyze the renormalization of the lattice theory, enumerating marginal operators consistent with the lattice symmetries. Thanks to the super-renormalizability of the theory, the counter-terms that need to be fine-tuned in order to restore full supersymmetry in the continuum limit can be fixed perturbatively at one loop. Since our lattice model was based on a complexification of 3d $\mathcal{N}=4$ SYM, we also propose to add mass terms in order to decouple the unwanted fields in the continuum limit.     

The study of 3d $\mathcal{N}=4$ lattice gauge theories is still in its infancy, so there are many important directions to explore:
\begin{itemize}
\item Perhaps the most immediate task is to analyze perturbative renormalization of the lattice theory along the lines of \cite{Catterall:2011pd}. With these fine-tunings in hand, we will then be in a position to simulate the model on a computer and check the predictions of \cite{Seiberg:1996nz}.

\item It would be interesting to generalize our lattice model to incorporate matter multiplets along the lines of \cite{Joseph:2013jya,Catterall:2015tta} in order to investigate 3d mirror symmetry, whereby two different 3d $\mathcal{N} = 4$ gauge theories flow to the same superconformal fixed point in the IR \cite{Intriligator:1996ex}. Note that under mirror symmetry, Wilson loops are exchanged with vortex loops \cite{ Assel:2015oxa}. Since lattice gauge theory is well-suited for the computation of loop operators, it should provide a powerful tool for testing such 3d dualities. 

\item Note that our construction can be generalized to $d=2$ by taking only two basis vectors of the lattice to be independent. It would therefore be interesting to study relation to the lattice formulations proposed in \cite{Cohen:2003qw,Sugino:2003yb} and invetigate 2d mirror symmetry \cite{Hori:2000kt}.

\item Using the gauge-invariant Hamiltonian formulation of Yang-Mills-Chern-Simons theories with $0 \leq \mathcal{N} \leq 4$ supersymmetry, it has been argued that a mass gap is present for $\mathcal{N} \leq 1$ and absent for extended supersymmetry \cite{Karabali:1995ps,Agarwal:2012bn}. It would therefore be interesting to explore how to formulate non-abelian Chern-Simons theory on a lattice and couple it to our model in order to test these arguments and explore the existence of a mass gap. Note that a lattice formulation of abelian Chern-Simons theory was proposed in \cite{Sen:2000ez}.

\item Holographic duals of 3d $\mathcal{N}=4$ superconformal field theories were proposed in \cite{Assel:2011xz}. Given that such theories arise as the IR fixed points of 3d $\mathcal{N}=4$ gauge theory, it would be very interseting to use lattice techniques to test these proposals. Morover, it would be interesting to use lattice techniques to  simulate certain 3d Euclidean Yang-Mills theories theories coupled to scalars and fermions which provide a holographic description of inflationary cosmology \cite{ McFadden:2010na}.
\end{itemize}

Ultimately, we hope that this work will provide a useful starting point for studying the non-pertubative dynamics of 3d $\mathcal{N}=4$ gauge theories using lattice techniques, as well as many other important questions in quantum field theory and holography.  

\begin{center}
\textbf{Acknowledgments}
\end{center}

We thank Simon Catterall, Stefano Cremonesi, Masanori Hanada, Anosh Joseph, and  Loganayagam R 
for useful conversations. AL is supported by the Royal Society as a Royal Society University Research Fellowship holder.  JG was supported in part by the U.S.~Department of Energy, Office of Science, Office of High Energy Physics,
Grant No.~DE-SC0013496.  Both authors express appreciation for NORDITA hosting the workshop "Holography
and Dualities 2016:  New Advances in String and Gauge Theory" where this work was
initiated. AL also benefited from discussions at the International Centre for Theoretical Sciences (ICTS) during the program "Nonperturbative and Numerical Approaches to Quantum Gravity, String Theory and Holography" (Code: ICTS/NUMSTRINGS/2018/01).

\appendix

\section{R-symmetry} \label{Rsymm}

In this Appendix, we analyze the discrete R-symmetries of the Donaldson-Witten twist of 4d $\mathcal{N}=2$ SYM. There are seven discrete R-charges. Four of them can be combined into
a 1-form $R_{\mu}$ and the remaining three into a self-dual 2-form
$R_{\mu\nu}=-R_{\nu\mu}=\star R_{\mu\nu}$. We will
determine the discrete R-symmetries and deduce the parameter $\alpha$
in the twisted Lagrangian in \eqref{4dfull} following the approach in \cite{Catterall:2013roa}. The other seven
supersymmetries can then be obtained by conjugating the nilpotent
supersymmetry by the seven discrete R-symmetries. For convenience, we reproduce the Lagrangian in \eqref{4dfull} below:
\[
g^{2}\mathcal{L}_{4d}^{\mathcal{N}=2}=\frac{1}{4}\mathcal{F}_{\mu\nu}\mathcal{F}^{\mu\nu}+\frac{1}{2}\mathcal{D}_{\mu}\bar{\phi}\mathcal{D}^{\mu}\phi-\alpha\left[\phi,\bar{\phi}\right]^{2}
\]

\begin{equation}
-i\chi^{\mu\nu}\mathcal{D}_{\mu}\psi_{\nu}-\frac{i}{2}\eta\mathcal{D}_{\mu}\psi^{\mu}-\frac{i}{2}\bar{\phi}\left\{ \psi_{\mu},\psi^{\mu}\right\} +i\alpha\phi\left\{ \eta,\eta\right\} +\frac{i}{8}\phi\left\{ \chi_{\mu\nu},\chi^{\mu\nu}\right\} .\label{eq:l4c2}
\end{equation}

\subsection{$R_{\mu}$}

Let us make the following ansatz for the transformations generated
by $R_{\mu}$:

\[
\eta\rightarrow\beta_{1}\psi_{\mu}
\]
\[
\psi_{\mu}\rightarrow\beta_{1}^{-1}\eta
\]
\[
\psi_{\nu}\rightarrow\beta_{2}\chi_{\mu\nu}
\]
\[
\chi_{\mu\nu}\rightarrow\beta_{2}^{-1}\psi_{\nu}
\]
\[
\chi_{\nu\rho}\rightarrow\beta_{2}^{-1}\epsilon_{\mu\nu\rho\lambda}\psi_{\lambda}
\]
\[
\phi\rightarrow\bar{\phi}
\]

\[
\bar{\phi}\rightarrow\phi
\]
\[
\mathcal{A}\rightarrow\mathcal{A}
\]
where $\nu,\rho\neq\mu$. Demanding that the Lagrangian in \eqref{eq:l4c2}
is invariant under these transformations fixes the coefficients to
be
\[
\alpha=\frac{1}{8},\,\,\,\beta_{1}=\pm2 i,\,\,\,\beta_{2}=\pm i.
\]

\subsection{$R_{\mu\nu}$}

Let us make the following ansatz for the transformations generated
by $R_{\mu\nu}$:

\[
\eta\rightarrow\gamma_{1}\chi_{\mu\nu}
\]
\[
\chi_{\mu\nu}\rightarrow-\gamma_{1}^{-1}\eta
\]
\[
\psi_{\mu}\rightarrow\gamma_{2}\psi_{\nu}
\]
\[
\psi_{\nu}\rightarrow-\gamma_{2}^{-1}\psi_{\mu}
\]
\[
\psi_{\rho}\rightarrow\epsilon_{\mu\nu\rho\lambda}\psi^{\lambda}
\]
\[
\chi_{\mu\rho}\rightarrow\gamma_{3}\chi_{\nu\rho}
\]
\[
\chi_{\nu\rho}\rightarrow-\gamma_{3}^{-1}\chi_{\mu\rho}
\]
\[
\chi_{\rho\lambda}\rightarrow-\gamma_{1}^{-1}\epsilon_{\rho\lambda\mu\nu}\eta
\]
\[
\left\{ \mathcal{A},\phi,\bar{\phi}\right\} \rightarrow\left\{ \mathcal{A},\phi,\bar{\phi}\right\} 
\]
where $\rho,\lambda\neq\mu,\nu$. Demanding that \eqref{eq:l4c2} is
invariant under the above transformations fixes the coefficients to
be

\[
\alpha=\frac{1}{8},\,\,\,\gamma_{1}=2,\,\,\,\gamma_{2}=\gamma_{3}=1.
\]

\section{Comparison of Twists} \label{comparisonblau}
In this appendix, we will compare the two twists of 3d $\mathcal{N}=4$
SYM. Recall that prior to twisting, the fermions and scalars transform
in the $\left(\frac{1}{2},\frac{1}{2},\frac{1}{2}\right)$ and $(0,1,0)$
representation of the global symmetry group $SU(2)_{E}\times SU(2)_{N}\times SU(2)_{R}$,
respectively, where $SU(2)_{E}$ is the group of Euclidean rotations,
$SU(2)_{N}$ is an internal symmetry group that has a 6d origin, and
$SU(2)_{R}$ is the R-symmetry group. Twisting by the R-symmetry group
amounts to breaking $SU(2)_{E}\times SU(2)_{R}\rightarrow SU(2)'={\rm diag}\left(SU(2)_{E}\times SU(2)_{R}\right)$,
after which the fields transform in the following representations of
$SU(2)'\times SU(2)_{N}$: 
\[
{\rm fermions}:\,\left(0,\frac{1}{2}\right)\oplus\left(1,\frac{1}{2}\right),\,\,\,{\rm bosons}:\,(1,0)\oplus(0,1).
\]
Note that this can be obtained from dimensionally reducing the Donaldson-Witten
twist of 4d $\mathcal{N}=2$ SYM, which is our starting point for defining the lattice
theory. On the other hand, it is also possible to twist by $SU(2)_{N}$
as shown by Blau-Thompson in \cite{Blau:1996bx}. Moreover, a lattice theory based on
this twist was developed in \cite{Joseph:2013jya}. Twisting by $SU(2)_{N}$ breaks $SU(2)_{E}\times SU(2)_{N}\rightarrow SU(2)'={\rm diag}\left(SU(2)_{E}\times SU(2)_{N}\right)$
after which the fields transform in the following representations
of $SU(2)'\times SU(2)_{R}$: 
\[
{\rm fermions}:\,\left(0,\frac{1}{2}\right)\oplus\left(1,\frac{1}{2}\right),\,\,\,{\rm bosons}:\,(1,0)\oplus(1,0).
\]
Note that in the Blau-Thompson twist, there are no bosonic scalars.

In the remainder of this appendix, we will derive the Lagrangians
for the two twists of 3d $\mathcal{N}=4$ SYM by dimensionally reducing
6d $\mathcal{N}=1$ SYM, demonstrating that they indeed describe the
same underlying theory even though their Lagrangians take different
forms and realize BRST symmetry in different ways (whereas the BRST
charge is nilpotent in the Blau-Thompson twist, it squares to a gauge transformation
in the Donaldson-Witten twist). We will also show that the 3d Donaldson-Witten twist can be obtained by dimensionally reducing
the Lagrangian in equation \eqref{4dfull}. 

The Lagrangian for 6d $\mathcal{N}=1$ SYM is given by 

\begin{equation}
\mathcal{L}_{6d}={\rm tr}\left(\frac{1}{4}F_{MN}F^{MN}- i \Psi_{R}^{\dagger}\Gamma^{M}D_{M}\Psi_{L}\right)\label{eq:6dlagrangian}
\end{equation}
where $M=1,...,6$, $D_{M}X=\partial_{M}X+\left[A_{M},X\right]$,
and $F_{MN}=\left[D_{M},D_{N}\right]$. Since we are working in Euclidean
signature, the Dirac matrices obey 
\[
\left\{ \Gamma_{M},\Gamma_{N}\right\} =\delta_{MN}.
\]
Morover$\Psi_{L/R}$ are chiral spinors satisfying 

\[
\Gamma_{7}\Psi_{L/R}=\pm\Psi_{L/R}
\]
where $\Gamma_{7}=\Gamma_{1}...\Gamma_{6}$. In the following we will
present explicit formulas for the Dirac matrices as tensor products of the Pauli
matrices: 
\[
\sigma_{0}=\left(\begin{array}{cc}
1 & 0\\
0 & 1
\end{array}\right),\,\,\,\sigma_{1}=\left(\begin{array}{cc}
0 & 1\\
1 & 0
\end{array}\right),\,\,\,\sigma_{2}=\left(\begin{array}{cc}
0 & -i\\
i & 0
\end{array}\right),\,\,\,\sigma_{3}=\left(\begin{array}{cc}
1 & 0\\
0 & -1
\end{array}\right).
\]

\subsection*{Blau-Thompson Twist}

Our derivation of the Blau-Thompson twist of 3d $\mathcal{N}=4$ SYM
will closely follow the original derivation in \cite{Blau:1996bx}. Consider the following
representation of the Dirac matrices which naturally split 6d into
$3+3$, making the $SU(2)_{E}\times SU(2)_{N}$ symmetry manifest:

\[
\Gamma_{k=1,2,3}=\sigma_{1}\otimes\sigma_{0}\otimes\sigma_{k},\,\,\,\Gamma_{a=4,5,6}=\sigma_{2}\otimes\sigma_{a-3}\otimes\sigma_{0},\,\,\,\Gamma_{7}=\sigma_{3}\otimes\sigma_{0}\otimes\sigma_{0}.
\]
In particular, $SU(2)_{E}$ is generated by $\frac{1}{2}\left[\Gamma_{k},\Gamma_{l}\right]$
with $k,l\in\left\{ 1,2,3\right\} $, and $SU(2)_{N}$ is generated by
$\frac{1}{2}\left[\Gamma_{a},\Gamma_{b}\right]$ with $a,b\in\left\{ 3,4,5\right\} $.
With this choice of Dirac matrices, the fermions can be written as
follows: 

\[
\Psi_{L}=\left(\begin{array}{c}
\psi^{\alpha\dot{\alpha}}\\
0
\end{array}\right),\,\,\,\Psi_{R}^{\dagger}=\left(\begin{array}{cc}
0 & \chi^{\alpha\dot{\alpha}}\end{array}\right),
\]
where $\alpha,\dot{\alpha}$ are spinor indices for $SU(2)_{E}\times SU(2)_{N}$.
To dimensionally reduce to three dimensions, we take the fields
to be independent of $x^{4},x^{5},x^{6}$, after which the covariant
derivatives along the internal directions reduce to 
\[
D_{a}X\rightarrow\left[A_{a},X\right].
\]
If we then twist by $SU(2)_{N}$, this breaks $SU(2)_{E}\times SU(2)_{N}\rightarrow{\rm diag}\left(SU(2)_{E}\times SU(2)_{N}\right)$
which amounts to identifying the $\alpha$ and $\dot{\alpha}$ indices.
It is then convenient to decompose the fermions into spin-1 and spin-0
parts as follows:
\[
\psi^{\alpha\beta}=\psi^{k}\left(\sigma_{k}\right)^{\alpha\beta}+\epsilon^{\alpha\beta}\eta,\,\,\,\chi^{\alpha\beta}=\chi^{\mu}\left(\sigma_{k}\right)^{\alpha\beta}+\epsilon^{\alpha\beta}\bar{\eta}.
\]
After doing so, the fermionic terms in \eqref{eq:6dlagrangian}
reduce to 

\begin{equation}
\mathcal{L}_{3d}^{f}={\rm tr}\left(-2\chi^{kl}\mathcal{D}_{k}\psi_{l}-2i\chi^{k}\bar{\mathcal{D}}_{k}\eta+2i\bar{\eta}\bar{\mathcal{D}}_{k}\psi^{k}\right)\label{eq:3dbtf}
\end{equation}
where we have defined $\chi_{kl}=\epsilon_{klm}\chi^{m}$. Morover,
after dimenional reduction and twisting by $SU(2)_{N}$, the internal components of $A_{M}$
become a vector in three dimensions:
\[
A_{M}=\left(A_{k},A_{a}\right)\rightarrow\left(A_{k},V_{k}\right).
\]
The bosonic terms in \eqref{eq:6dlagrangian} then reduce to 

\[
\mathcal{L}_{3d}^{b}={\rm tr}\left(\frac{1}{4}F_{kl}^{2}+\frac{1}{4}\left[V_{k},V_{l}\right]^{2}+\frac{1}{2}\left(D_{k}V_{l}\right)^{2}\right)={\rm tr}\left(\frac{1}{4}\left(F_{kl}-\left[V_{k},V_{l}\right]\right)^{2}+\frac{1}{4}\left(D_{[k}V_{l]}\right)^{2}+\frac{1}{2}\left(D_{k}V^{k}\right)^{2}\right),
\]
where we used integration by parts to obtain the second equality.
It it then convenient to define the following covariant derivatives:

\[
\mathcal{\mathcal{D}}_{k}X=\partial_{k}X+\left[\left(A+iV\right)_{k},X\right],\,\,\,\bar{\mathcal{D}}_{k}X=\partial_{k}X+\left[\left(A-iV\right)_{k},X\right],
\]
in terms of which the bosonic terms in the Lagrangian can be written
compactly as follows: 

\begin{equation}
\mathcal{L}_{3d}^{b}={\rm tr}\left(\frac{1}{4}\mathcal{\bar{F}}_{kl}\mathcal{F}^{kl}+\frac{1}{2}\left(D_{k}V^{k}\right)^{2}\right)\label{eq:3dbtb},
\end{equation}
where $\mathcal{F}_{kl}=\left[\mathcal{D}_{k},\mathcal{D}_{l}\right]$ and
$\bar{\mathcal{F}}_{kl}=\left[\bar{\mathcal{D}}_{k},\bar{\mathcal{D}}_{l}\right]$.

\subsection*{Donaldson-Witten Twist}

Our strategy for deriving the Donaldson-Witten twist of 3d $\mathcal{N}=4$
SYM will be to first dimensionally reduce the 6d Lagrangian in \eqref{eq:6dlagrangian}
to 4d to make $SU(2)_{R}$ manifest, and then dimensionally reduce
to three dimensions and twist by $SU(2)_{R}$. We therefore choose
the following representation for the Dirac matrices which make the
$4+2$ split manifest:
\[
\Gamma^{\mu}=\sigma_{0}\otimes\sigma_{1}\otimes\sigma^{\mu-1},\,\,\,\Gamma^{5}=\sigma_{1}\otimes\sigma_{3}\otimes\sigma_{0},\,\,\,\Gamma^{6}=\sigma_{2}\otimes\sigma_{3}\otimes\sigma_{0},\,\,\,\Gamma^{7}=-\sigma_{3}\otimes\sigma_{3}\otimes\sigma_{0}
\]
where $\mu=1,2,3,4$. Note that $\frac{1}{2}\left[\Gamma_{\mu},\Gamma_{\nu}\right]$
generate the 4d Euclidean rotation group which is locally $SU(2)_{l}\times SU(2)_{r}$.
For this choice of Dirac matrices, the fermions can be written as
\[
\Psi_{L}=\left(\begin{array}{c}
0\\
\lambda^{\alpha}\\
\bar{\omega}^{\dot{\alpha}}\\
0
\end{array}\right),\,\,\,\Psi_{R}^{\dagger}=\left(\begin{array}{cccc}
\bar{\lambda}^{\dot{\alpha}} & 0 & 0 & \omega^{\alpha}\end{array}\right),
\]
where $\alpha,\dot{\alpha}$ are spinor indices for $SU(2)_{l}\times SU(2)_{r}$.
If we take the fields to be independent of the $x^{5},x^{6}$ directions,
\eqref{eq:6dlagrangian} reduces to the following 4d Lagrangian: 

\begin{equation}
\mathcal{L}_{4d}={\rm tr}\left(\frac{1}{4}F_{\mu\nu}F^{\mu\nu}+\frac{1}{2}D_{\mu}\phi D^{\mu}\bar{\phi}-\frac{1}{8}\left[\phi,\bar{\phi}\right]^{2}-i\bar{\psi}^{I\dot{\alpha}}D_{\alpha\dot{\alpha}}\psi_{I}^{\alpha}-\frac{i}{2}\left\{ \psi^{I\alpha},\psi_{I\alpha}\right\} \bar{\phi}+\frac{i}{2}\left\{ \bar{\psi}^{I\dot{\alpha}},\bar{\psi}_{I\dot{\alpha}}\right\} \phi\right),\label{eq:dw4d}
\end{equation}
where $\phi=A_{5}-iA_{6}$, $\bar{\phi}=\phi^{\dagger}$, $D_{\alpha\dot{\alpha}}=\frac{1}{2} \sigma_{\alpha\dot{\alpha}}^{\mu} D_{\mu}$,
and $I$ is an $SU(2)$ R-symmetry index; in particular the fermionic
fields are given by 
\[
\psi^{I\alpha}=\sqrt{2}\left(\lambda^{\alpha},\omega^{\alpha}\right),\,\,\,\bar{\psi}^{I\dot{\alpha}}=\sqrt{2}\left(\bar{\lambda}^{\dot{\alpha}},\bar{\omega}^{\dot{\alpha}}\right).
\]
Next, we dimensionally reduce to three dimensions by taking the fields
to be independent of $x^{4}$. Relabelling the fields as $\left(\phi,\bar{\phi}\right)=\left(B_{1}+iB_{2},B_{1}-iB_{2}\right)$
and $A^{4}=B_{3}$, the bosonic terms in the Lagrangian reduce to
\begin{equation}
\mathcal{L}_{3d}^{b}={\rm tr}\left(\frac{1}{4}F_{kl}^{2}+\frac{1}{2}\sum_{i=1}^{3}\left(D_{k}B_{i}\right)^{2}+\frac{1}{2}\sum_{i<j}\left[B_{i},B_{j}\right]^{2}\right).\label{eq:dw3db}
\end{equation}
Note that dimensional reduction to 3d breaks $SU(2)_{l}\times SU(2)_{r}\rightarrow SU(2)_{E}={\rm diag}\left(SU(2)_{l}\times SU(2)_{r}\right)$,
which amounts to identifying the $\alpha$ and $\dot{\alpha}$ indices. If we then
twist the 3d rotation group by $SU(2)_{R}$, this breaks $SU(2)_{E}\times SU(2)_{R}\rightarrow{\rm diag}\left(SU(2)_{E}\times SU(2)_{R}\right)$
so the R-symmetry index $I$ can be identified with the $SU(2)_E$ index $\alpha$. It
is then convenient to decompose the fermions into spin-1 and spin-0
parts as follows: 
\[
\psi^{\alpha\beta}=\psi^{k}\left(\sigma_{k}\right)^{\alpha\beta}+\epsilon^{\alpha\beta}\psi^{4},\,\,\,\bar{\psi}^{\alpha\beta}=\chi^{k}\left(\sigma_{k}\right)^{\alpha\beta}+\frac{1}{2}\eta\epsilon^{\alpha\beta}.
\]
After doing so, the fermionic terms in \eqref{eq:dw4d} become 

\[
\mathcal{L}_{3d}^{f}={\rm tr}\left(-i\chi^{kl}D_{k}\psi_{l}-i\chi^{k}D_{k}\psi^{4}+i\chi^{k}\left[B_{3},\psi_{k}\right]-\frac{i}{2}\psi^{k}D_{k}\eta-\frac{i}{2}\psi^{4}\left[B_{3},\eta\right]\right.
\]
\begin{equation}
\left.+\frac{i}{8}\phi\left\{ \eta,\eta\right\} -\frac{i}{2}\bar{\phi}\left\{ \psi^{k},\psi_{k}\right\} -\frac{i}{2}\bar{\phi}\left\{ \psi^{4},\psi_{4}\right\} +\frac{i}{2}\phi\left\{ \chi^{k},\chi_{k}\right\} \right) \label{eq:dw3df},
\end{equation}
where we defined $\chi_{kl}=\epsilon_{klm}\chi^{m}$. The Lagrangian
in equations \eqref{eq:dw3db} and \eqref{eq:dw3df} is precisely what we obtain
after dimensionally reducing the twisted 4d Lagrangian in \eqref{4dfull}.

\section{Link field potential}
\label{lfp}
Here we delve into details of the link field potential \myref{lfpeq}, which
is of the same form as is used in 4d $\Ncal=4$ SYM, and thus well understood
from those previous studies.  To further understand this functional, we use the lattice spacing $a$ to
rescale the link fields to have canonical dimension 1, so that the
potential then takes the form
\beq
{\cal V} = m_U^2 \sum_{\mu,x} \( \frac{1}{N} \tr \cUb_\mu \cU_\mu - \frac{1}{a^2} \)^2
\eeq
For this we explore the continuum limit in the linear formulation
\beq
\cU_\mu(x) = \frac{1}{a} + A_\mu(x) + i B_\mu(x), \quad \cUb_\mu(x) = \frac{1}{a} - A_\mu(x) + i B_\mu(x)
\label{thecont}
\eeq
Here, $B_\mu$ are scalar fields that must be lifted to very large masses
in order to obtain the target continuum theory, 
whereas $A_\mu$ contain the physical gauge fields.
In addition, one linear combination of these also contains a third scalar field (in addition
to $\phi, {\bar \phi}$) that should be retained
in the low energy spectrum,
\beq
\phi_3 = \frac{1}{2} \sum_{\mu=1}^4 A_\mu = \sum_{\mu=1}^4 P_{4\mu} A_\mu
\eeq
The latter notation indicates a projection.  The actual gauge fields $V_i$ are obtained from
orthogonal projections:
\beq
V_i = \sum_{\mu=1}^4 P_{i \mu} A_\mu, \quad i = 1, 2, 3
\eeq
where $\sum_\mu P_{\alpha \mu} P_{\beta \mu} = \delta_{\alpha \beta}$; $\alpha, \beta = 1, \ldots, 4$.

Multiplying everything out and tracing, throwing out the gauge fields $V_\mu$, we obtain
the quantity 
\beq
\tr \cU_\mu \cUb_\mu = \frac{N}{a^2} - \frac{2 \sqrt{N}}{a} B_a^0 + \frac{1}{4} \phi_3^A \phi_3^A + B_\mu^A B_\mu^A
\label{tUUb}
\eeq
where $A=0,1,\ldots,N^2-1$ correspond to the U(N) generators $t^A$ which
satisfy $\tr t^A t^B = -\delta^{AB}$.  Note in particular that the U(1) generator is
\beq
t^0 = \frac{i}{\sqrt{N}}
\eeq
Thus the scalar potential is
\beq
{\cal V} = m_U^2 \sum_{\mu,x} \left[ \frac{4}{a^2 N} (B_\mu^0)^2 - \frac{4 B_\mu^0}{a N \sqrt{N}} 
\( \frac{1}{4} \phi_3^A \phi_3^A
        + B_\mu^A B_\mu^A \) 
+ \frac{1}{N^2} \( \frac{1}{4} \phi_3^A \phi_3^A
+ B_\mu^A B_\mu^A \)^2 \right]
\label{stabpot}
\eeq
What we see is the following:  the 4 U(1) scalars in $B_\mu^0$ get a positive mass-squared term, driving them to zero
at small field values.  This is good because they would shift the lattice spacing if they had a nonzero vacuum
expectation value.  We see that
at large field values where the quartic term dominates, there are no flat directions.  This
is highly desireable, because we do not want scalars wandering along flat
directions during a simulation, but rather stuck near a point in moduli space.  All six U(N) scalars are lifted and
there is no runaway.  At intermediate fields the cubic term plays an important role and there will be saddle points.
At first the potential decreases as $\sum_\mu B_\mu^0$ is increased, but then finally the quartic term takes over to prevent
runaway.

Next notice what happens in the limit of very small lattice spacing.  The $B_\mu^0$ mass term
is dominant and lifts these modes from the low-energy spectrum.  Also, these fields are
driven to zero in the semi-classical approximation.  The subdominant (in lattice spacing)
cubic term is therefore driven to zero, and is negligible.  If we scale $m_U^2$ (dimensionless)
to smaller values as we decrease $a$, such that $m_U^2/a^2$ is nevertheless very large,
then the quartic term is also negligible.  In this way the non-mass potential terms do
not disturb the scalar potential of the desired low energy theory.

\bibliographystyle{JHEP}
\bibliography{3dN=4bib}
\end{document}